\documentclass[3p,times,procedia]{elsarticle}

\usepackage{ecrc}

\volume{00}

\firstpage{1}

\journalname{Physics Procedia}

\runauth{M.~Grefe and T.~Delahaye}

\jid{phpro}

\jnltitlelogo{Physics Procedia}

\CopyrightLine{2014}{Published by Elsevier Ltd.}

\usepackage{amssymb}
\usepackage{amsmath}

\usepackage[colorlinks=true]{hyperref}

\begin{document}

\begin{frontmatter}



\dochead{TAUP 2013 \\ 13th International Conference on Topics in Astroparticle and Underground Physics}

\title{Antiproton Limits on Decaying Gravitino Dark Matter}

\author[label1]{Michael Grefe\corref{cor1}}
\ead{michael.grefe@desy.de}
\address[label1]{Instituto de F\'isica Te\'orica UAM/CSIC and Departamento de F\'isica Te\'orica, \\[-0.5mm] Universidad Aut\'onoma de Madrid, Cantoblanco, E-28049 Madrid, Spain}
\cortext[cor1]{Corresponding author}
\author[label2]{Timur Delahaye}
\address[label2]{The Oskar Klein Centre for Cosmoparticle Physics, Department of Physics, \\[-0.5mm] Stockholm University, AlbaNova University Center, SE-106 91 Stockholm, Sweden}



\begin{abstract}
We report on constraints on the lifetime of decaying gravitino dark matter in models with bilinear \textit{R}-parity violation derived from observations of cosmic-ray antiprotons with the PAMELA experiment. Performing a scan over a viable set of cosmic-ray propagation parameters we find lower limits ranging from $8\times 10^{28}\,$s to $6\times 10^{28}\,$s for gravitino masses from roughly 100\,GeV to 10\,TeV. Comparing these limits to constraints derived from gamma-ray and neutrino observations we conclude that the presented antiproton limits are currently the strongest and most robust limits on the gravitino lifetime in the considered mass range. These constraints correspond to upper limits on the size of the bilinear \textit{R}-parity breaking parameter in the range of $10^{-8}$ to $8\times 10^{-13}$.
\end{abstract}

\begin{keyword}
dark matter decay \sep gravitino dark matter \sep cosmic-ray antiprotons
\PACS 95.35.+d \sep 96.50.S- \\
\textit{Report numbers:} FTUAM-14-40 \sep IFT-UAM/CSIC-14-002
\end{keyword}

\end{frontmatter}



\section{Introduction}
\label{introduction} 
The gravitino in models with bilinear $R$-parity violation is an attractive dark matter (DM) candidate since this type of models provides a possibility to reconcile the generation of the baryon asymmetry in the Universe via thermal leptogenesis with the constraints on late-decaying particles from big bang nucleosynthesis (BBN)~\cite{Buchmuller:2007ui}: The high reheating temperature required for thermal leptogenesis naturally leads to a thermally produced gravitino relic abundance that matches the observed DM density in the Universe if the gravitino mass is in the range of roughly ten to several hundreds of GeV. A tiny violation of $R$-parity is sufficient to trigger the decay of the next-to-lightest supersymmetric particle (NLSP) before the time of BBN, thus maintaining the successful predictions of standard cosmology. The double suppression of the gravitino decay width by the Planck scale and the small amount of $R$-parity violation leads to a gravitino lifetime largely exceeding the age of the Universe, thus making it a viable DM candidate~\cite{Takayama:2000uz}. 

An intriguing feature of this type of models is that decaying gravitino DM exhibits a rich phenomenology although the gravitino usually is thought to be one of the most elusive DM candidates. Interestingly, the cosmological constraints on the size of $R$-parity violation allow for a gravitino lifetime that can be probed by cosmic-ray experiments, \textit{i.e.} the decay of gravitino DM in the Galactic halo can lead to observable signals in the spectra of cosmic rays. In this proceedings article we report on lower limits on the gravitino lifetime and the corresponding upper limits on the amount of $R$-parity violation that were derived from observations of cosmic-ray antiprotons with the PAMELA experiment~\cite{Adriani:2012paa,Adriani:2010rc}. We also compare the obtained lifetime limits to constraints derived from gamma-ray and neutrino observations. For a more detailed discussion of the presented results we refer the interested reader to~\cite{Delahaye:2013yqa}.
 \vspace{-2pt}

\section{Gravitino Decay}
\label{decay}
 \vspace{-1pt}
 
\begin{figure}[ht]
 \centering
 \includegraphics[width=0.325\textwidth]{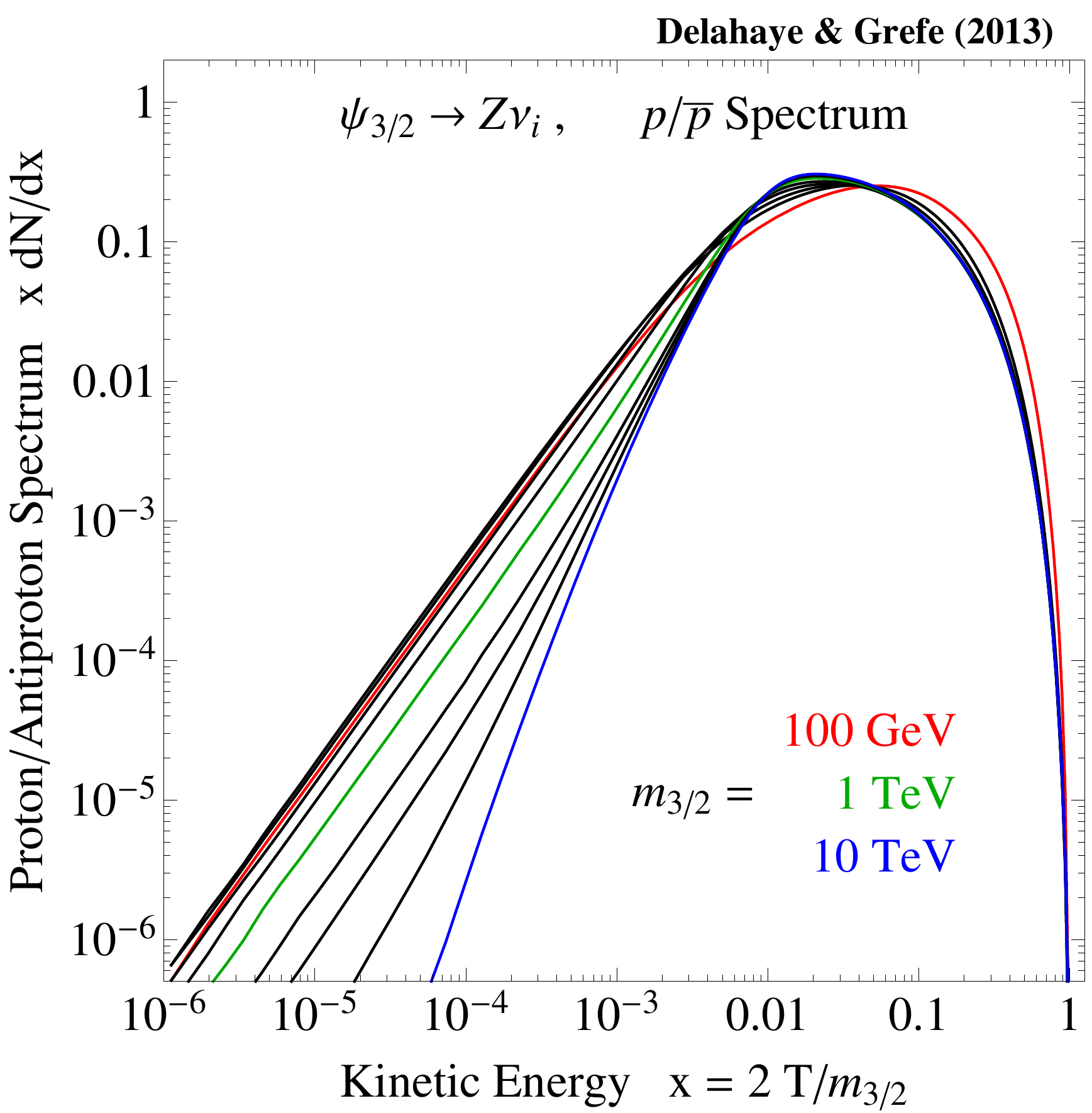} 
 \hfill
 \includegraphics[width=0.325\textwidth]{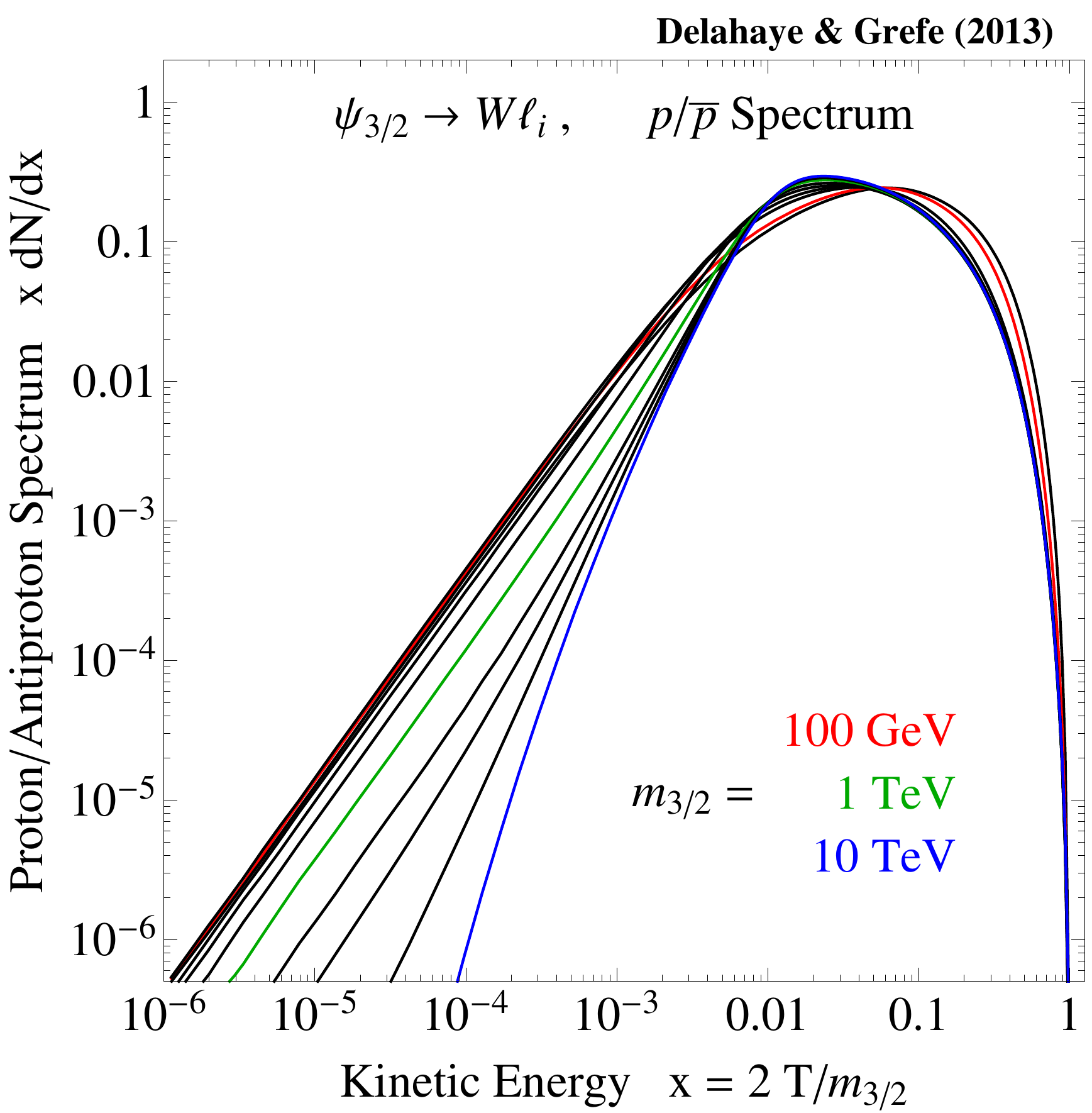} 
 \hfill
 \includegraphics[width=0.325\textwidth]{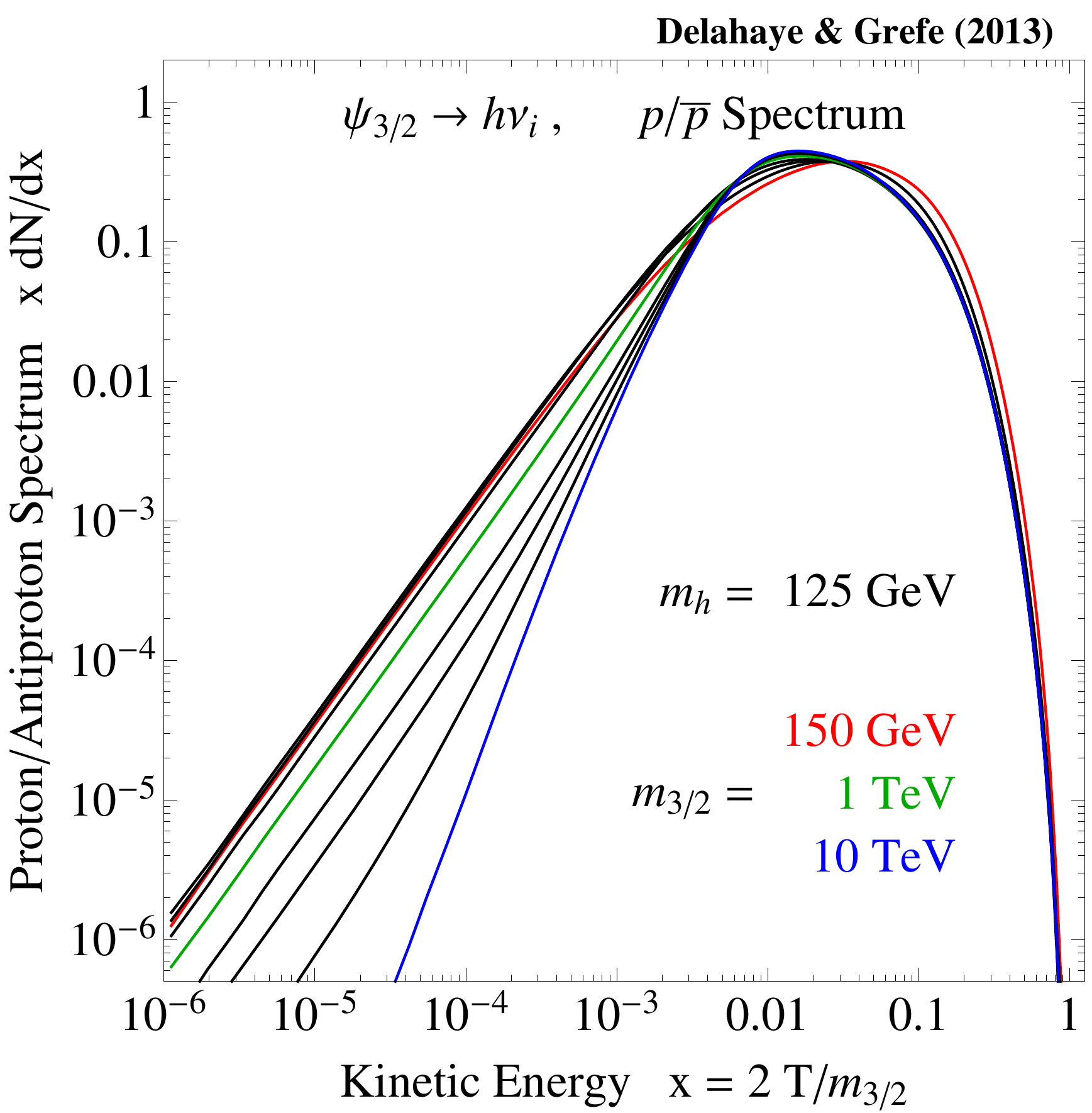} 
 \caption{Proton/antiproton spectra from the decay of the gravitino into $Z\nu_i$ (\textit{left}), $W\ell_i$ (\textit{centre}) and $h\nu_i$ (\textit{right}). The spectra are shown for gravitino masses of 85\,GeV (only $W\ell_i$) 100\,GeV ($W\ell_i$ and $Z\nu_i$), 150\,GeV, 200\,GeV, 300\,GeV, 500\,GeV, 1\,TeV, 2\,TeV, 3\,TeV, 5\,TeV, and 10\,TeV. They are normalized to the respective gravitino mass to allow for a common presentation of the spectra for all masses.}
 \label{spectra}
\end{figure}

The gravitino can decay into several two-body final states via bilinear $R$-parity breaking interactions: $\psi_{3/2}\rightarrow\gamma\nu_i, Z\nu_i, W\ell_i$ and $h\nu_i$. Depending on the gravitino mass only a subgroup of these final states may be kinematically accessible for on-shell production. Since the decay channel $\gamma\nu_i$ at leading order does not produce any antiprotons in the final state, we concentrate on the channels that include massive bosons and therefore on gravitino masses above the $W$ boson mass. The spectra of protons and antiprotons from these decay channels were obtained by a simulation with the event generator \textsc{Pythia} 6.4~\cite{Sjostrand:2006za} and the results are presented in Fig.~\ref{spectra} for a set of gravitino masses $m_{3/2}$ ranging from 85\,GeV to 10\,TeV.
 \vspace{-5pt}

\begin{figure}[ht]
\centering
  \includegraphics[width=0.504\linewidth]{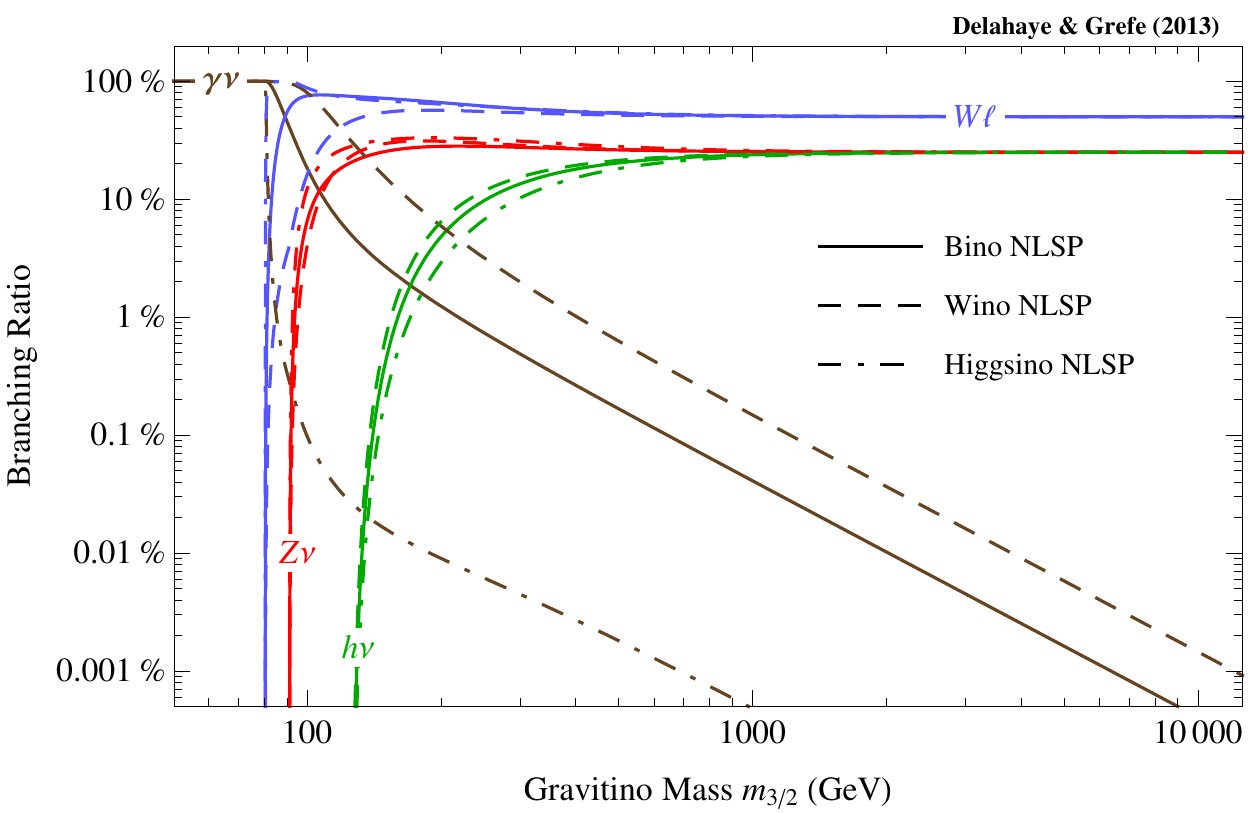} 
  \hfill
  \includegraphics[width=0.49\linewidth]{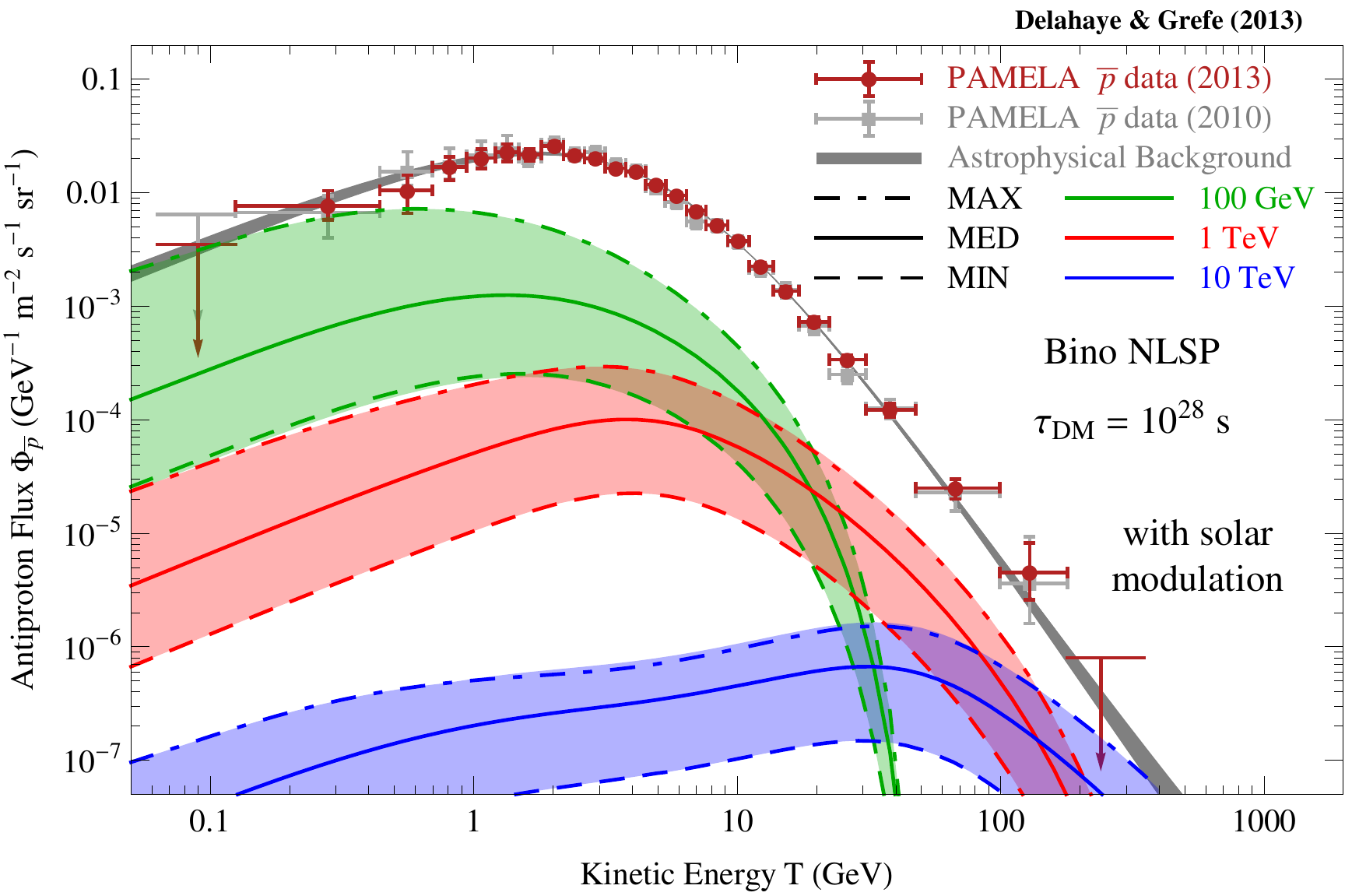}
  \caption{\textit{Left:} Branching ratios of the different gravitino two-body decay channels as a function of the gravitino mass for three choices of supersymmetry parameters: Bino NLSP, Wino NLSP and Higgsino NLSP. \textit{Right:} Antiproton flux expected from gravitino decay compared to the expected flux of antiprotons from secondary production and data from PAMELA. The gravitino signal is calculated for the case of Bino NLSP, a lifetime of $10^{28}$\,s and for three different masses: 100\,GeV, 1\,TeV and 10\,TeV. It is shown for MIN/MED/MAX propagation parameters and the coloured bands also show the allowed range of spectra from the scan over propagation parameters.}
  \label{fig:BRplot-gravitinoantiproton}
\end{figure}

The antiproton spectrum from gravitino decay is a combination of the above decay channels according to their respective branching ratios. In the left panel of Fig.~\ref{fig:BRplot-gravitinoantiproton} we show the branching ratios of the different gravitino decay channels as a function of the gravitino mass. We considered three example sets of supersymmetry parameters presenting different cases for the neutralino NLSP: a case with a Bino-like NLSP, a case with a Wino-like NLSP and a case with a Higgsino-like NLSP. The branching ratios of the $Z\nu_i, W\ell_i$ and $h\nu_i$ channels for the different cases of supersymmetry parameters only differ significantly for gravitino masses between the $W$ boson mass and a few hundred GeV. The effect on the $\gamma\nu_i$ decay channel is more significant, but in any case its branching ratio drops quickly for gravitino masses above the $W$ boson mass. For the discussion in these proceedings we focus on the case of gravitino DM with a Bino-like NLSP.
 \vspace{-2pt}

\section{Antiproton Limits on the Gravitino Lifetime}
\label{lifetime}
 \vspace{-1pt}

Decays of gravitino DM in the Milky Way halo produce antiprotons that subsequently propagate through the Galaxy and eventually might give a visible contribution to the antiproton spectrum as measured by cosmic-ray experiments. In fact, antiprotons, being roughly a factor of $10^4$ less abundant than protons in the cosmic radiation, are an ideal channel to search for exotic sources of cosmic rays like DM decays. This is because antiprotons as opposed to protons are not produced in astrophysical sources but only by secondary production in spallation of high-energy cosmic rays on the interstellar medium.
\newpage

In the right panel of Fig.~\ref{fig:BRplot-gravitinoantiproton} we compare the expected antiproton flux from gravitino decays to the data of the PAMELA experiment~\cite{Adriani:2012paa,Adriani:2010rc} and the flux of secondary antiprotons. The data are well explained by secondary antiprotons and there is no apparent need for any DM contribution. Therefore, antiproton observations allow to set strong constraints on the gravitino lifetime.

The flux of secondaries is well constrained by the observation of other cosmic-ray species. In particular the boron-to-carbon ratio plays an important role in constraining the parameters describing cosmic-ray propagation through the Milky Way. Thus the flux uncertainty is on the level of the experimental error bars. By contrast, the flux coming from DM decays is subject to larger uncertainties since decays happen in the spherical halo while secondary antiprotons, like other cosmic-ray species that are used to constrain the propagation parameters, have their source in the Galactic plane only. The resulting uncertainty amounts to roughly one order of magnitude. On a more local scale the flux of cosmic rays is influenced by the solar wind. We treated this so-called solar modulation with the Fisk potential method. 

\begin{figure}[t!]
\centering
  \includegraphics[width=0.488\linewidth]{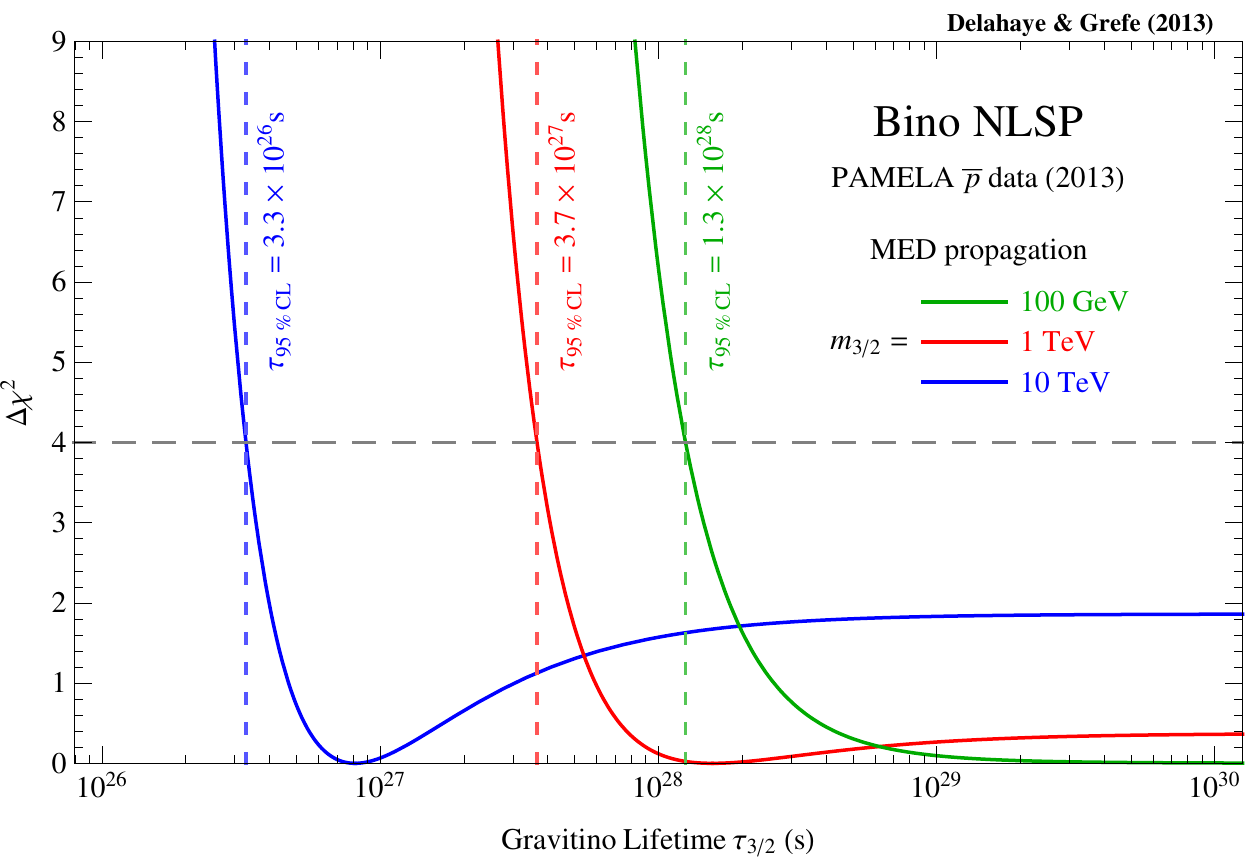}
  \hfill
  \includegraphics[width=0.503\linewidth]{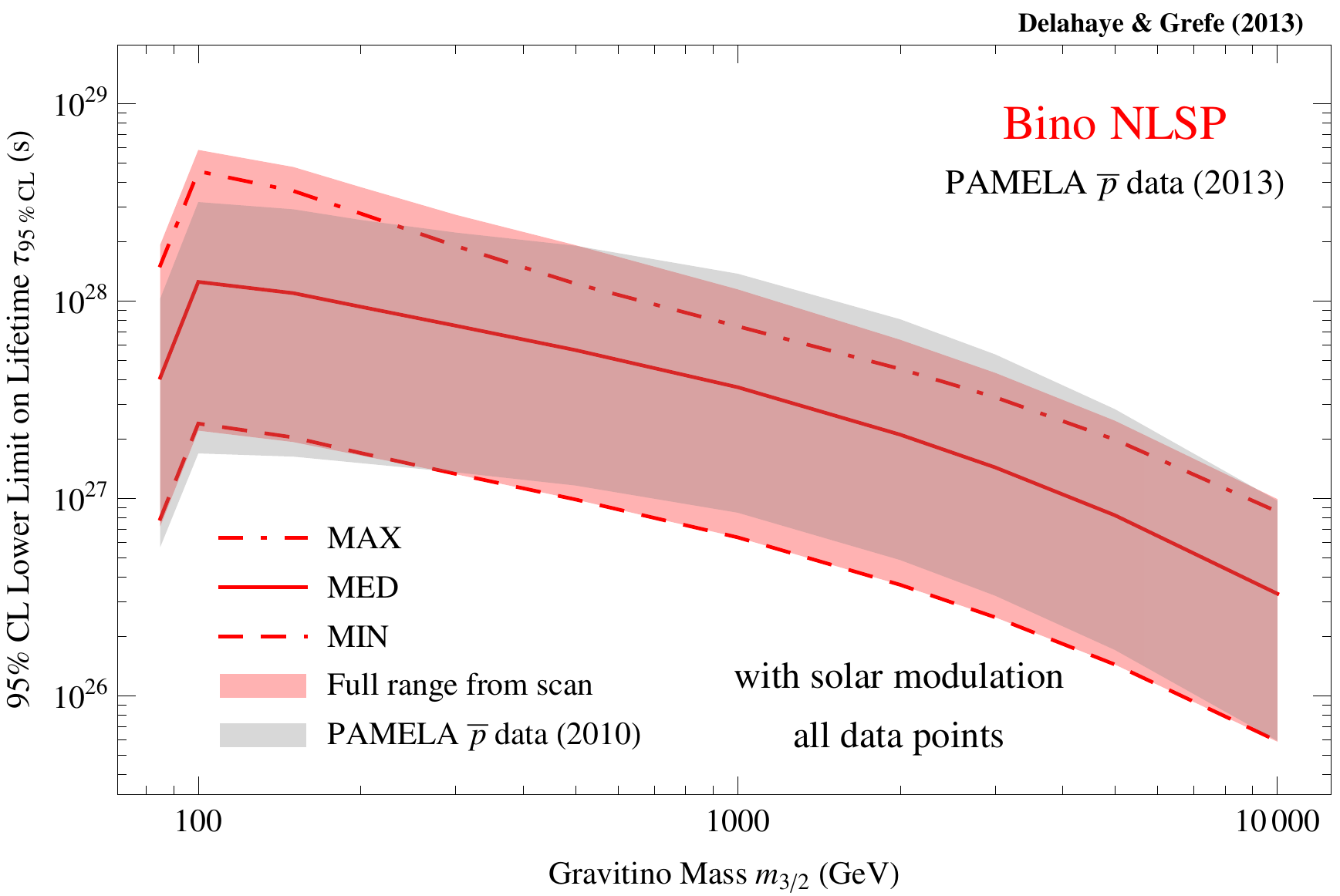}
  \caption{\textit{Left:} $\Delta\chi^2$ as a function of the gravitino lifetime $\tau_{3/2}$ for the case of gravitino DM with a Bino-like NLSP and MED cosmic-ray propagation parameters. We show three example cases for the gravitino mass: 100\,GeV (green), 1\,TeV (red) and 10\,TeV (blue). We set lower limits on the gravitino lifetime at $\Delta\chi^2=4$, corresponding to $95\%$ CL. \textit{Right:} Constraints on the gravitino lifetime derived from antiproton observations by the PAMELA experiment. We show the constraints for the case of Bino NLSP. Gravitino lifetimes below the dashed/solid/dash-dotted line are excluded at 95\,\% CL for MIN/MED/MAX propagation parameters. In addition, the coloured band shows the whole range of propagation uncertainty derived from a scan over allowed propagation parameters. We also present the range of limits derived from a previous PAMELA data release.}
  \label{fig:deltachi2-gravitinolifetime}
\end{figure}

We calculated lower limits on the gravitino lifetime at 95\,\% CL by requiring that the limiting $\chi^2(\tau_{95\,\%\:\mathrm{CL}})$ deviates from the best-fit $\chi^2(\tau_{\mathrm{best\:fit}})$ by an amount $\Delta\chi^2$ corresponding to a $2\sigma$ exclusion:
\begin{equation}
  \chi^2(\tau_{95\,\%\:\mathrm{CL}})=\chi^2(\tau_{\mathrm{best\:fit}})+\Delta\chi^2\,. \nonumber
\end{equation}
In the left panel of Fig.~\ref{fig:deltachi2-gravitinolifetime} we show an example for $\Delta\chi^2$ as a function of the gravitino lifetime for three gravitino masses and MED propagation parameters from~\cite{Donato:2003xg}. For very large lifetimes gravitino decays do not contribute at all to the $\chi^2$ statistics. In several cases -- in particular for larger gravitino masses -- we found a best-fit lifetime that slightly improves the fit because the central values of the PAMELA data points are slightly above the expected flux of secondary antiprotons. However, the deviation lies well within the error bars of the data and this plot confirms that there is no strong statistical significance of a non-vanishing DM contribution. Therefore, in the right panel of Fig.~\ref{fig:deltachi2-gravitinolifetime} we only present lower limits on the gravitino lifetime.

Three sources of uncertainty currently limit improvements of the constraints presented here. The first comes from the rather poor knowledge of the cosmic-ray propagation parameters and we sized this uncertainty in this work. A novelty of this analysis is to not only use a predefined set of propagation parameters (MIN/MED/MAX) but to scan over a large set of parameters that are compatible with cosmic-ray data~\cite{Maurin:2001sj}. This is possible because we treated antiproton propagation through the Milky Way by a CPU-friendly method, namely by means of semi-analytical solutions of the cosmic-ray diffusion equation~\cite{Bringmann:2006im}.

Another important source of uncertainty are the spectra of primary cosmic rays like protons and $\alpha$ particles. Although these species are abundant and quite well measured, there are some discrepancies among the results of different cosmic-ray experiments. A third source of uncertainty is the limited knowledge of nuclear cross sections that enter into the calculation of antiproton production in cosmic-ray spallation.

A strong hope for significant improvements in this field lies in the AMS-02 experiment on the International Space Station. A thorough study of data on all possible cosmic-ray species from a single experiment would allow to narrow down the uncertainty on the propagation parameters. In addition, AMS-02 will be able to increase the observed energy range of antiprotons. Therefore, AMS-02 data could lead to strongly improved constraints or even discover a signal from gravitino decay.

\section{Comparison with other Cosmic-Ray Constraints}
\label{other} 
\begin{figure}[t]
\centering
  \includegraphics[width=0.49\linewidth]{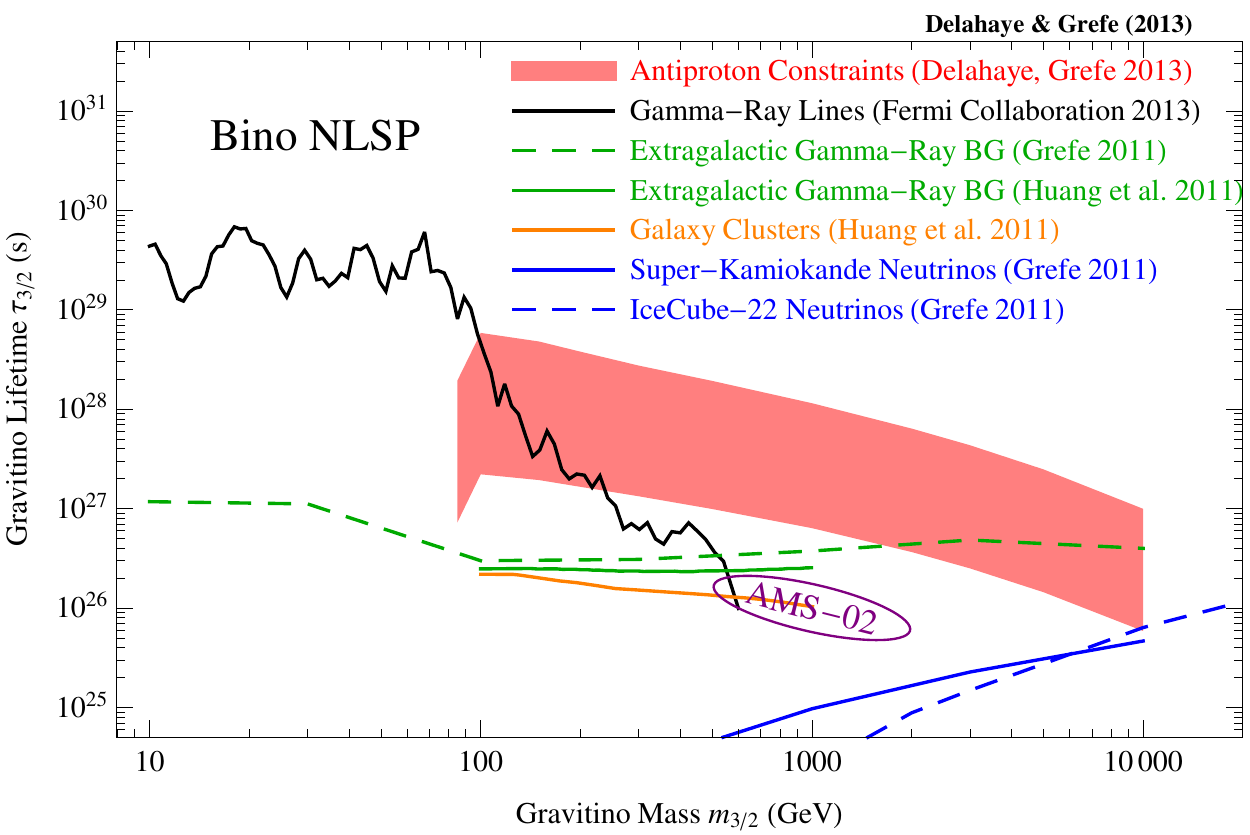}
  \hfill
  \includegraphics[width=0.495\linewidth]{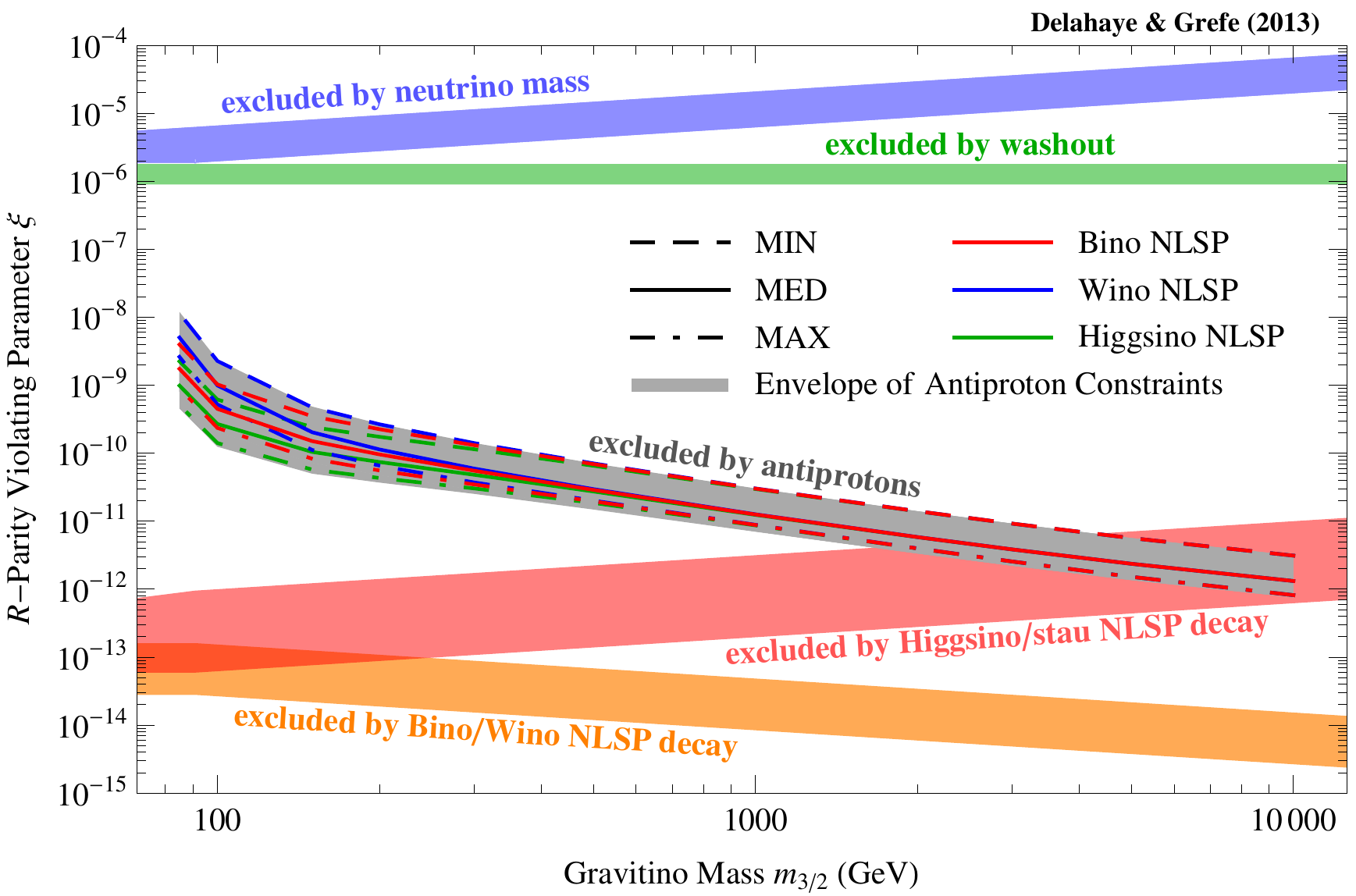}
  \caption{\textit{Left:} Comparison of the lower limits on the gravitino lifetime derived from antiproton data with limits from gamma-ray line searches, the extragalactic gamma-ray background, gamma-rays from galaxy clusters, and neutrinos. The area below the lines is excluded. For comparison we show in purple the gravitino parameters needed to explain the positron fraction as observed by AMS-02. \textit{Right:} Constraints on the amount of $R$-parity violation as a function of the gravitino mass. We compare the upper limits derived from antiproton data to upper limits from contributions to the neutrino mass and from the washout of the baryon asymmetry in the early Universe, and to lower limits from BBN constraints on the late decay of different NLSP candidates.}
  \label{fig:other-xibound}
\end{figure}

In addition to the antiproton limits discussed above, also the non-observation of a gravitino decay signal in other cosmic-ray channels can be used to constrain the gravitino lifetime. A compilation of various lifetime limits discussed in the literature is presented in the left panel of Fig.~\ref{fig:other-xibound}.\footnote{Note that these lifetime limits are not all set at the same confidence level. The aim of this figure is just to give a qualitative idea of how the constraining potential of different cosmic-ray channels compares with each other.} For light gravitinos the strongest limits come from gamma-ray line searches~\cite{Fermi-LAT:2013uma}. Above the $W$ boson mass threshold the significance of these limits is weakened by the steeply falling branching ratio of the $\gamma\nu_i$ decay channel (the steepness of the falloff is quite model-dependent, \textit{cf.} Fig.~\ref{fig:BRplot-gravitinoantiproton}). Additional gamma-ray constraints were derived in~\cite{Grefe:2011dp} and~\cite{Huang:2011xr} by comparing the diffuse flux of prompt gamma rays from gravitino decays in the Galactic halo and at extragalactic distances to the Fermi-LAT determination of the extragalactic gamma-ray background (EGB)~\cite{Abdo:2010nz},\footnote{The authors of~\cite{Huang:2011xr} used slightly more conservative assumptions to derive the limits, but since their results are only available for a narrower range of masses, we present the results from~\cite{Grefe:2011dp} for comparison.} and in~\cite{Huang:2011xr} from the comparison of the expected contribution of gravitino decays in galaxy clusters to the observed gamma-ray spectra of several clusters. For very heavy gravitinos with masses above 10\,TeV also neutrino bounds become significant. The presented bounds were derived in~\cite{Grefe:2011dp} from a neutrino flux limit from the direction of the Galactic centre presented by the Super-Kamiokande collaboration~\cite{Desai:2004pq} and from a search for neutrino lines from DM decay in the Galactic halo with the IceCube 22-string configuration~\cite{Abbasi:2011eq}. We conclude that despite the large propagation uncertainty the antiproton limits derived in this work are currently the strongest and most robust constraints on the gravitino lifetime for gravitino masses from roughly 100\,GeV to 10\,TeV.

However, the gamma-ray and neutrino channels have great potential for future improvements: Updated Fermi-LAT data on the EGB extending to higher energies and a proper modelling of astrophysical contributions to the EGB are expected to improve the presented gamma-ray constraints notably. In addition, a much larger neutrino data set from the complete IceCube detector together with a dedicated analysis of gravitino decays as currently undertaken by the IceCube collaboration\footnote{J.~Pepper, C.~P\'erez~de~los~Heros, and C.~Rott, private communication (2013).} is expected to lead to a significant improvement on the neutrino constraints.

The authors of~\cite{Ibe:2013nka} studied gravitino decay in the context of the AMS-02 data on the positron fraction~\cite{Aguilar:2013qda} and found that a gravitino with a mass of roughly 1\,TeV and a lifetime of roughly $10^{26}\,$s could explain the observed rise above 10\,GeV.\footnote{The region displayed in Fig.~\ref{fig:other-xibound} is not a confidence region but is just meant to indicate the approximate parameter values needed to explain the data with an inclination following the expected mass dependence for DM decay.} However, this conclusion is in strong tension with constraints on the associated production of antiprotons and gamma rays. Therefore we conclude that gravitino decays cannot contribute significantly to the AMS-02 data and an astrophysical explanation of the observed rise in the positron fraction is required in this scenario.

\section{Constraints on the Strength of \boldmath$R$-parity Violation}
\label{rparity} 
The lower limits on the gravitino lifetime can also be used to derive upper limits on the size of the bilinear \textit{R}-parity breaking parameter $\xi\propto \tau_{3/2}^{-1/2}$. The resulting constraints are presented in the right panel of Fig.~\ref{fig:other-xibound} together with several constraints coming from other considerations: Bilinear \textit{R}-parity violation generates a contribution to the neutrino masses and therefore the upper limit on the sum of neutrino masses from cosmological observations, $\sum m_{\nu_i}<0.23\,$eV~\cite{Ade:2013zuv}, leads to an upper bound on $\xi$. Another upper limit comes from the requirement that a primordially generated baryon asymmetry must not be washed out by \textit{R}-parity breaking interactions before the electroweak phase transition~\cite{Endo:2009cv}. In addition to that, lower bounds on the size of \textit{R}-parity violation can be derived from the fact that the NLSP has to decay before the time of BBN in order not to spoil the successful predictions of the light element abundances. This bound depends on the nature of the NLSP.

We conclude that for the considered gravitino mass range from roughly 100\,GeV to 10\,TeV the constraints coming from antiproton observations are several orders of magnitude stronger than those derived from the contribution to the neutrino masses and the washout of the baryon asymmetry. However, they are not strong enough to completely exclude the parameter space together with lower limits from late NLSP decay. Only for specific choices of the NLSP and gravitino masses beyond 1\,TeV there is no viable parameter space remaining for this model. More generally, an improvement of the antiproton limits on $\xi$ of roughly three orders of magnitude will be necessary to robustly rule out the model in the considered mass range. That corresponds to an improvement of the lifetime limits by six orders of magnitude. In view of the uncertainties connected to an indirect search in a background-dominated channel like antiprotons, it appears highly unlikely that this goal can be achieved in the foreseeable future. Nevertheless, any improvement on these searches is more than welcome and might also lead to a detection of a signal from gravitino DM decay.

\section{Conclusions}
\label{conclusions}
We presented strong limits on the lifetime of unstable gravitino dark matter and the strength of bilinear $R$-parity violation derived from PAMELA antiproton data. Although gravitino decays produce cosmic-ray spectra that could explain the PAMELA excess in the positron fraction, this possibility is in strong tension with constraints from the associated production of antiprotons and gamma rays. It is also expected that neutrino telescopes like IceCube will soon have the capability to provide competitive constraints on heavy gravitinos. In conclusion, we find that for gravitino masses above the $W$ mass threshold antiprotons currently provide the strongest and most robust constraints on the gravitino lifetime. The upcoming data of the AMS-02 experiment will even improve the sensitivity of this channel.
 \vspace{-1pt}

\section*{Acknowledgements}
The work of MG is supported by the Marie Curie ITN "UNILHC" under grant number PITN-GA-2009-237920, by the Spanish MINECO’s “Centro de Excelencia Severo Ochoa” Programme under grant SEV-2012-0249 and by the Comunidad de Madrid under the grant HEPHACOS S2009/ESP-1473. The work of TD is supported by the Swedish Research Council under contract number 349-2007-8709.
 \vspace{-1pt}








\end{document}